\documentclass[sigconf]{acmart}
\acmConference[ASEW 2022]{The 37th IEEE/ACM International Conference on Automated Software Engineering Workshops}{10 - 14 October, 2022}{United States}

\usepackage{multirow}
\usepackage{balance}

\AtBeginDocument{%
  \providecommand\BibTeX{{%
    \normalfont B\kern-0.5em{\scshape i\kern-0.25em b}\kern-0.8em\TeX}}}

\copyrightyear{2022}
\acmYear{2022}
\setcopyright{acmcopyright}\acmConference[ASE '22]{37th IEEE/ACM International Conference on Automated Software Engineering}{October 10--14, 2022}{Rochester, MI, USA}
\acmBooktitle{37th IEEE/ACM International Conference on Automated Software Engineering (ASE '22), October 10--14, 2022, Rochester, MI, USA}
\acmPrice{15.00}
\acmDOI{10.1145/3551349.3560438}
\acmISBN{978-1-4503-9475-8/22/10}

\begin{document}


\title[How Readable is Model-generated Code?]{How Readable is Model-generated Code? \\ Examining Readability and Visual Inspection of GitHub Copilot}

\author{Naser Al Madi}
\email{nsalmadi@colby.edu}
\orcid{0000-0003-4551-3080}
\affiliation{%
  \institution{Colby College}
  \city{Waterville}
  \state{Maine}
  \country{USA}
}

\renewcommand{\shortauthors}{Al Madi et al.}

\begin{abstract}

\textbf{Background:} Recent advancements in large language models have motivated the practical use of such models in code generation and program synthesis. However, little is known about the effects of such tools on code readability and visual attention in practice.
\\
\textbf{Objective:} In this paper, we focus on GitHub Copilot to address the issues of readability and visual inspection of model generated code. Readability and low complexity are vital aspects of good source code, and visual inspection of generated code is important in light of automation bias.
\\
\textbf{Method:} Through a human experiment (n=21) we compare model generated code to code written completely by human programmers.  We use a combination of static code analysis and human annotators to assess code readability, and we use eye tracking to assess the visual inspection of code.
\\
\textbf{Results:} Our results suggest that model generated code is comparable in complexity and readability to code written by human pair programmers.  At the same time, eye tracking data suggests, to a statistically significant level, that programmers direct less visual attention to model generated code.
\\
\textbf{Conclusion:} Our findings highlight that reading code is more important than ever, and programmers should beware of complacency and automation bias with model generated code.

\end{abstract}



\maketitle
\section{Introduction}

Artificial Intelligence (AI) has been used in software development in general and in generating code in particular since the 1970's \cite{barenkamp2020applications, manna1971toward}. Recent advances in Large Language Models, such as GPT-2 \cite{radford2019language} and GPT-3 \cite{brown2020language}, have motivated the use of such models in code generation and program synthesis \cite{chen2021evaluating, li2022competition}. GitHub Copilot\footnote{https://copilot.github.com/} is a software development tool that offers code generation of lines, code chunks, or even entire programs based on existing code and comments \cite{sobania2021choose}. Copilot is built with the GPT-3 language model \cite{chen2021evaluating} and trained with large sets of public code.  Copilot is marketed as an AI powered pair-programmer, a software development practice where two programmers collaboratively write a single piece of code.

Pair programming matches two programmers with similar or different levels of experience to work together on writing code. Programmers periodically switch between two roles, driver and navigator. The driver controls the mouse and keyboard and writes code while the navigator observes the driver's work and critically thinks about defects, structural issues, and alternative solutions, while looking at the larger picture of the code \cite{Adam, williams2001integrating}. This practice has been shown to increase code quality and enjoyment during the problem solving process \cite{williams2001integrating}.  

Several studies \cite{Vaithilingam2022expectation, sobania2021choose, pearce2021can, ziegler2022productivity} have examined aspects of productivity, usability, and security of Copilot generated code. However, little is known about the effects of such tools on code readability and visual attention in practice. In addition to the importance of generating readable code, automation bias could lead some programmers to accept suggested code without inspection, resulting in defects that are discovered at a latent stage of the software development process, where fixing defects is significantly more costly.

Therefore, in this paper we aim to study GitHub Copilot empirically with eye tracking in a natural software development environment (VS Code IDE). We focus on the issues of code readability and visual inspection of Copilot generated code. Through an experiment with 21 participants we compare code written with Copilot to code written by human pair-programmers (as driver and navigator). Using static code analysis, human readability annotators, and eye tracking we aim to answer the following research questions:

\begin{itemize}

\item \textit{RQ1: How readable is Copilot generated code compared to code written completely by human programmers?}

\item \textit{RQ2: How well do programmers inspect Copilot generated code?}

\end{itemize}

Our results have implications to Copilot users and creators, highlighting the importance of reading and inspecting model generated code and being cautious of automation bias.

\section{Experiment}
In this section, we present the details of our \textbf{IRB approved} experiment and the sources of data we used to answer our research questions. Our within-subject study consists of 21 participants in total, with data from 15 participants used in the eye tracking analysis. We make the our data and replication package publicly available through the following Open Science Framework \href{https://osf.io/ghfqz/}{[LINK].}

\subsection{Participants}
The 21 participants in our study were Computer Science students with experience ranging from first year to fourth year.  Some participants are in their first Computer Science course (CS1), while other students have completed several internships as professional programmers.  The programming experience of participants ranged from 1 to 8 years, and 12 out of 21 participants were familiar with pair-programming. All participants were above the age of 18, and 8 were Female, 12 Male, and 1 Non-binary.  Participation was voluntary, and participants were awarded a $\$10$ gift card after the experiment.   Also, some eye tracking recordings for participants who wear corrective glasses were corrupt or had error beyond 1.5 degrees, therefore they were dropped from the eye tracking data analysis (their code is still used in the code analyses). Therefore, the eye tracking analysis is based on data from 15 participants.

\subsection{Apparatus}
The experiment lab consisted of a pair-programming station with two screens, two keyboards, and two mice connected to a single computer.  The eye movement of the participant is tracked with an EyeLink 1000 Plus (SR Research) eye tracker, with a sampling frequency of 1000 samples per second. The screen resolution was set to 1920x1080.  The development environment is Visual Studio Code (VS Code) with Copilot add-on that could be activated or deactivated between trials.

\subsection{Procedure}
First, the participant is entered into the experiment lab, and the general procedure of the experiment is explained along with an online consent form.  If the participant agrees to proceed with the experiment, the online form collects the participant's programming experience in years, demographics, and whether they had prior experience with pair-programming or not.  Regardless of the participant's experience with pair-programming, every participant watches a video explaining pair-programming and the roles of driver and navigator and how they typically alternate.

Participants worked on developing text-based Minesweeper game in Python. None of the participants had implemented this game before, and the participants were familiarized with the rules of the game by playing several rounds of the game prior to the recorded development task. After that, the participant's seat is adjusted for comfort along with a chin-rest that guarantees that the participant is captured by the eye tracker for the full duration of the experiment.  An eye tracker calibration and validation steps are taken to guarantee that eye movement is being tracked and recorded accurately on the computer screen. 

The development task was done under three conditions in randomized order. The conditions are pair programming with Copilot, pair programming  as a driver with another human volunteer, and pair programming as a navigator with another human volunteer. The time allocated is 20 minutes for Copilot, 10 minutes as a driver, and 10 minutes as a navigator (20 minutes with a human and 20 minutes with Copilot). The order of the three trial conditions was randomized to prevent the experience effect from influencing the results.  The experience effect describes how participants perform better after each trial as they gain experience with the task and environment.

\section{Methods}

\subsection{Static Code Analyses and Readability:}

Despite the widespread use of static code analyses metrics \cite{nunez2017source}, depending completely on such metrics can be problematic \cite{kaner2004software}.  Static code metrics are heuristics that often fail to measure what they aim to measure \cite{scalabrino2017automatically}.  Nonetheless, a recent study by Peitek et al. found that some source code metrics correlate with programmer cognition as measured by fMRI \cite{peitek2021program}.

Therefore, we cautiously combine static code metrics with readability assessment by human annotators to gain insights on the complexity and readability of source code.  To answer our research questions, we collect the following metrics from each trial using Radon\footnote{https://pypi.org/project/radon/} and Pycodestyle\footnote{https://pypi.org/project/pycodestyle/}:

\begin{itemize}
    \item Code Size: Logical Lines Of Code (LLOC).
        
    \item Complexity: Cyclomatic Complexity.
    
    \item Maintainability: Microsoft Visual Studio Maintainability.
    
    \item Vocabulary: Halstead Vocabulary.
    
    \item Volume: Halstead Volume.
    
    \item Difficulty: Halstead Difficulty.
    
    \item Coding style adherence: Number of coding style errors according to PEP8 coding style guide for Python \cite{van2001pep}.
    
\end{itemize}

We divide the number of coding style errors in each trial by the number of lines of code in the trial to calculate normalized coding style errors per line of code.  This normalization accounts for the variations in the number of lines of code added in each trial, without influence from how much code was written in the trial.  The idea here is that comparing the metrics of code written with Copilot to code written completely by human programmers, will allow us to see if Copilot code is more complex or substantially different from human written code.  

In addition, we ask five programmers (two female, and three male) with two to five years of programming experience to annotate the readability of 20 code snippets from our experiment through an online survey.  The code snippets are from Copilot, driver, and navigator trials, yet the origin of the code was hidden from the annotators.  The annotators were given general written guidelines with examples of readable and unreadable code, and they were asked to asses the readability of each snippet on the scale: very unreadable, unreadable, neutral, readable, very readable.  In addition, the annotators were asked to justify their readability assessment in writing.

\subsection{Eye Tracking:}

We developed our own tool for tracking eye movement in the VS Code IDE. Since existing tools for eye tracking in IDEs such as iTrace \cite{sharif2016itrace} do not yet support VS Code, and Copilot was only available on VS Code during the time of our experiment.

\begin{figure}[h]
    \centering
    \includegraphics[width=3.1in]{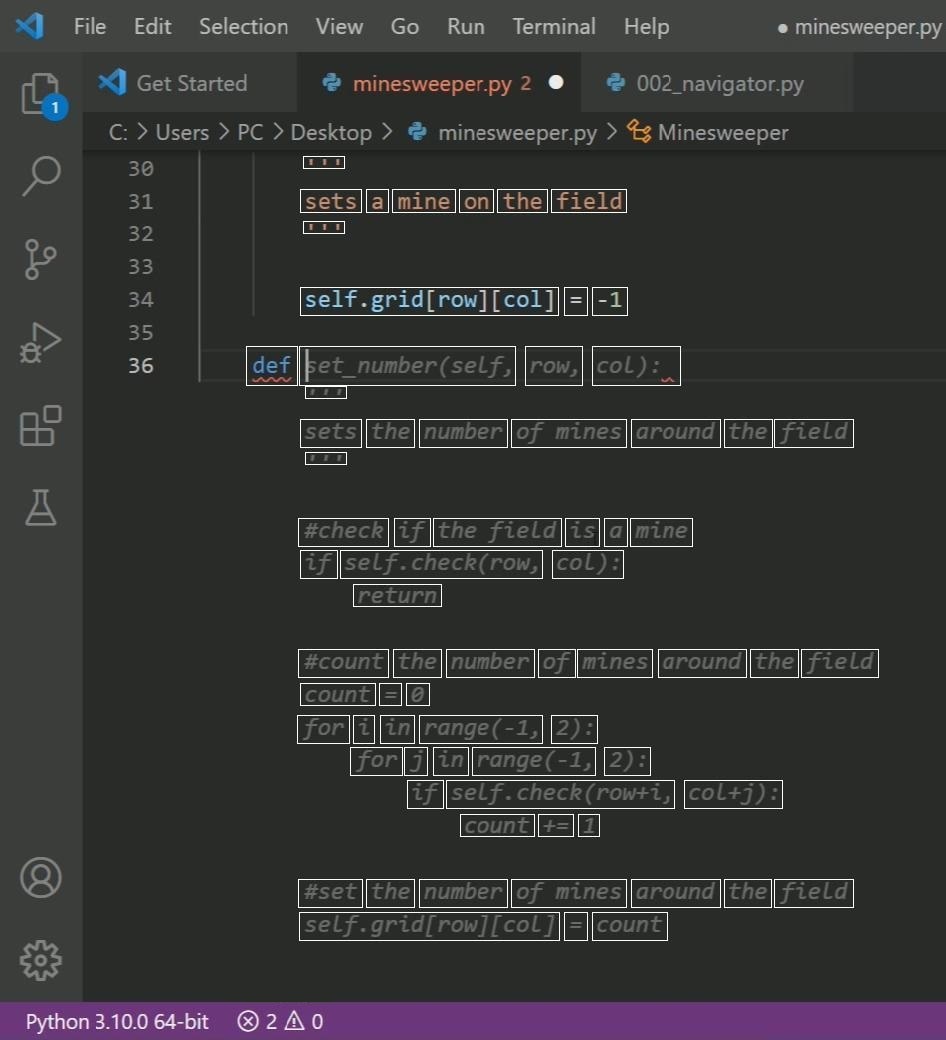}
    \caption{A sample program showing token level AOIs as generated by our tool for processing eye tracking data in VS Code.  Code in grey is suggested by Copilot.}
    \label{fig:rectangles}
\end{figure}

Our tool generates token-level Areas Of Interest (AOIs) as shown in Figure \ref{fig:rectangles}. With this, we can conduct a hit test to examine if a fixation was over a token or not using the Eye-Movement in Programming Toolkit (EMTK) \cite{al2021emip}.  This process is repeated until all the fixations in the experiment were analyzed. With this procedure, we were able to collect relevant eye movement data in a dynamic and natural coding environment (VS Code).

From fixation durations over AOIs we can calculate eye movement metrics that give us insights on the cognitive processes involved in reading code \cite{sharafi2020practical, al2021novice}. We use the following metrics to compare eye movement during the Copilot trial to driver and navigator trials:

\begin{itemize}

    \item Fixation Count: The total number of fixations on a token of code. 
    
    \item Total Time: The sum of all fixation durations on a token of code (in millisecond). 

    \item First-Fixation Duration: The duration of the first fixation on a token of code (in millisecond).  
    
    \item Single-Fixation Duration: The duration of the fixation when only one fixation was made on the token (in millisecond).
    
\end{itemize}

The connection between eye movement and cognition has been well established in cognitive psychology \cite{rayner1998eye}, where First Fixation Duration has been connected to lexical access (word identification), or the retrieval of word sound and meaning from memory \cite{rayner1998eye, reichle2003ez}. Total Time gives an indication of the time taken to complete lexical access and comprehension, where longer durations indicate more processing. These metrics have been used extensively in psychology and software engineering research \cite{sharafi2020practical}, with some of the earliest studies using fixation count and fixation duration to find the AOIs that attract more attention \cite{crosby1990we, crosby2002roles, uwano2006analyzing}. In summary, we associate fixation count and longer fixations with more cognitive processing.

\section{Results}

\subsection{Readability Analyses}
Our first research question (RQ1) focuses on the readability and complexity of GitHub Copilot generated code compared to code written by human programmers. To answer this question, we use static code analysis and a group of human annotators to assess the readability of code created in each condition (i.e., Copilot, Driver, Navigator).

\begin{table}[h]
\caption{Comparing mean static code metrics of code written in each trial condition (i.e., Copilot, Driver, and Navigator). $p$ shows the probability that Copilot, Driver, and Navigator data belong to the same population. Bold when \textit{$p$} \textless 0.01.}
\label{tab:metrics}
\begin{tabular}{ccccc}
                                                                 & \textbf{Copilot} & \textbf{Driver} & \textbf{Navigator} & \textit{\textbf{p}} \\ \hline
\begin{tabular}[c]{@{}c@{}}Code Size\\ (LLOC)\end{tabular}       & 19               & 11              & 15                 & .03                 \\ \hline
\begin{tabular}[c]{@{}c@{}}Cyclomatic \\ complexity\end{tabular} & 3                & 2               & 2                  & .06                 \\ \hline
Maintainability                                                  & 23               & 24              & 27                 & .96                 \\ \hline
\begin{tabular}[c]{@{}c@{}}Halstead\\ vocabulary\end{tabular}    & 13               & 9               & 12                 & .14                 \\ \hline
\begin{tabular}[c]{@{}c@{}}Halstead\\ volume\end{tabular}        & 222              & 149             & 142                & .08                 \\ \hline
\begin{tabular}[c]{@{}c@{}}Halstead\\ difficulty\end{tabular}    & 3                & 1               & 2                  & \textbf{.006}       \\ \hline
\begin{tabular}[c]{@{}c@{}}Style errors\\ / LLOC\end{tabular}    & 0.21             & 0.54            & 0.26               & .08                 \\ \hline
\end{tabular}
\end{table}

We normalize the logical lines of code (LLOC) by the duration of the trials, therefore our numbers correspond to equal durations of each condition.  Comparing mean code metrics in Table \ref{tab:metrics}, we see that programmers on average wrote slightly more code during the Copilot trial, in addition the Halstead volume of Copilot code appears larger than code written during Driver and Navigator trials.  Cyclomatic complexity, maintainability, and Halstead Vocabulary appears comparable between conditions.  

On the other hand, Halestead difficulty is higher with Copilot and the number coding style errors per line of code is higher in the driver trial.  Nonetheless, statistical tests show that only Halstead difficulty is statistically different between the three conditions. A Kruskal-Wallis test found statistically significant difference between the Halstead difficulty of Copilot, driver, and navigator codes (H(2) = 10.13, \textit{p} = .006).  Pairwise comparisons using Dunn's test indicated that Copilot Halstead difficulty scores were observed to be significantly different from those of the Driver code (p = .004). No other differences were statistically significant.  The effect size for this analysis (d = .99) was found to exceed Cohen’s convention for a large effect (d = .80). 

These results suggest that the code created during the Copilot trial is largely comparable to code written by human programmers during the driver and navigator trials.  At the same time, the results suggest that code written during the Copilot trial appears to be more difficult to comprehend according to Halstead difficulty metric, on a statistically significant level.

\begin{figure}[h]
    \centering
    \includegraphics[width=3.3in]{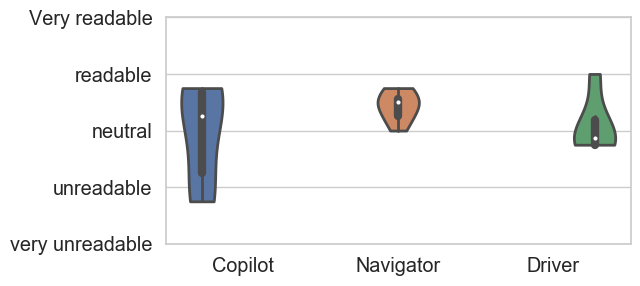}
    \caption{Code readability assessment by human annotators for code written during the Copilot, Driver, and Navigator trials.}
    \label{fig:readability_annotation}
\end{figure}

Furthermore, we approach RQ1 and code readability from the angle of human annotators who assessed the readability of code snippet that were written during Copilot, Driver, or Navigator trials. Figure \ref{fig:readability_annotation} shows a violin plot of readability assessment, where the white dot is the median, the upper and lower limits are the maximum and minimum values respectively, and the shape is a histogram of the data points. The figure shows comparable median values for readability, although some Copilot snippets received lower readability assessment by human annotators.

Using a scale between 0 and 4 to represent readability scores between "very unreadable" to "very readable," we find that the mean readability score for Copilot is 2.1, for navigator is 2.1, and for driver 2.4.  Nonetheless, A Kruskal-Wallis test found no statistically significant difference between the readability assessment for Copilot, driver, and navigator codes (H(2) = 1.52, \textit{p} = .46).

Overall, The readability analyses results suggest that Copilot generated code is comparable in complexity and readability to code written entirely by human programmers.  The only exception is Halstead difficulty metric, where copilot code appeared more difficult on a statistically significant level.  We will reflect more on these results in the discussion section.

\subsection{Eye Tracking Analyses}
In this section, we focus on the eye movement of programmers while programming with Copilot compared with programming with a human programmer.  Our focus in RQ2 is to examine how well do programmers inspect Copilot generated code? Our hypothesis is that code during a Copilot trial receives less scrutiny and inspection compared with code written with a human pair-programmer. Therefore, we examine Fixation Count, Total Time, First Fixation Duration, and Single Fixation Duration across trial conditions.

\begin{table}[h]
\caption{Comparing eye movement metrics of programmers in each trial condition (duration in milliseconds). $p$ shows the probability that Copilot, Driver, and Navigator data belong to the same population. Bold when \textit{$p$} \textless 0.01.}
\label{tab:beg-compare}
\begin{tabular}{lllll}
\hline
Metric (Mean)       & Copilot & Driver & Navigator & \textit{$p$} \\ \hline
Fixation Count      & 2.10    & 2.38   & 2.33      & \textbf{<.001} \\
Total Time          & 585     & 696    & 664       & \textbf{<.001} \\
First Fixation      & 273     & 284    & 279       & \textbf{<.001}\\
Single Fixation     & 282     & 292    & 281       & .032
\end{tabular}
\end{table}

Analyzing the eye movement of programmers in each condition, we compare eye movement metrics during Copilot, driver, and navigator trials in Table \ref{tab:beg-compare}.  The results show higher average fixation counts and higher fixation duration in the driver and navigator trials compared to the Copilot trial.

One-way ANOVA tests were performed to compare the effect of trial condition on eye movement metrics. The tests revealed that there was a statistically significant difference in Fixation Count, Total Time, and First Fixation duration between at least two groups (p<.001), as shown in Table \ref{tab:beg-compare}.  Tukey’s HSD Test for multiple comparisons found that the mean values of Fixation Count, Total Time, and First Fixation Duration were significantly different between Copilot and Driver (p  < .001), and Copilot and Navigator (p < .001).  There was no statistically significant difference in Single Fixation Duration between Copilot, Driver, and Navigator (p = .032). 

These results suggest that programmer's make fewer and shorter fixations while using Copilot, in comparison to pair-programming with a human.  These results suggest that programmers direct less visual attention to Copilot generated code.

\section{Discussion}

Our code analyses and readability annotation results in RQ1 suggest that code written with Copilot is comparable in complexity and readability to code written fully by human programmers. The only outlier to this is Halstead Difficulty, which was higher with Copilot, to a statistically significant level. These results match the results from Xu et al. which found no statistical difference in complexity between model generated code and code written by humans \cite{xu2022ide}.

On the other hand, our eye tracking results of RQ2 suggest that programmers make fewer fixations and spend less time reading code during the Copilot trial. This might be an indicator of less inspection or over-reliance on AI (automation bias), as we have observed some participants accept Copilot suggestions with little to no inspection.  This has been reported by another paper that studied Copilot \cite{Vaithilingam2022expectation}.  In our experiment, we also observed that participants sometimes interact with large chunks of suggested code during a Copilot trial, and sometimes the suggested code requires scrolling up and down to view fully.

Although the productivity benefits of Copilot remain questionable \cite{ziegler2022productivity}, our results suggest that reading and inspecting model-generated code is important.  Over-confidence in model generated code can lead to costly maintainability problems and issues in code quality.  For model creators, there are a number of ideas to aid code readability when generating code.  First, incorporating readability in the model itself to prioritize the most readable suggestion, if multiple suggestions are available. This requires new code readability metrics that match the readability assessment of human programmers. Second, generating large chunks of code all at once possibly hinders code inspection, perhaps presenting shorter code suggestions more gradually would aid code inspection. Third, programmers spend more time reading code than writing code, in fact from an economic perspective code comprehension takes the larges proportion of software development time and cost \cite{barry1981software}. Therefore applying large code models, such as Copilot Codex, to aid code comprehension could be very beneficial to the software engineering community.

\section{Threats to Validity}

\emph{Internal validity}: In regards to internal validity, it is typical for eye tracking studies to use static stimuli, and that does not match the real-life interactive experience of programming.  Therefore, we created a custom tool for eye tracking in the VS Code IDE, with an interactive experience that matches the activities of real-life software development. To mitigate threats of eye tracker accuracy, we used a high frequency eye tracker (1000 samples per second) with a high spatial resolution, and we corrected drift using a universal offset where all fixations are moved together. In addition, we dropped recordings that had higher error than 1.5 degrees, to make sure our eye tracking data is reliable.  

\emph{External validity}: A threat to the generalization of our results is that our participants were students. This was partially mitigated by the inclusion of students with widely varying degrees of expertise, ranging from 1 year to 8 years of programming experience, yet we plan to include industry professionals in a future study.

\emph{Construct validity}: Our code metrics were direct objective counts. The relationship between code metrics and cognition is still not established, therefore we cautiously use code metrics purely to compare Copilot code to human written code.  In addition, we utilize human annotation to get a "ground truth" of readability to cover the gap between static code metrics and readability. 

\emph{Conclusion validity}: Every statistical test has the threat of producing a false positive or negative, and we attempted to mitigate this threat by applying Bonferroni correction when we perform multiple-comparisons. Nonetheless, as with every empirical experiment, multiple studies are needed to verify results, which we encourage by making our data and code publicly available.

\section{Future Work}
We plan to recruit a larger sample with professional developers from the industry in our experiment.  In addition, we plan to include other human factors in the assessment of model-generated code. Eye tracking can also be used to assess cognitive load and other aspects of interaction between programmers and AI models.


\balance
\bibliographystyle{ACM-Reference-Format}
\bibliography{sample-base}

\end{document}